\documentclass{elsart}
\usepackage{natbib}
\usepackage{graphicx}
\usepackage{amssymb}

\begin{document}

\begin{frontmatter}


\title{AXPs/SGRs: Magnetars or Quark-stars?}

\author{Renxin Xu}
\address{School of Physics, Peking University, Beijing 100871,
China;\\
{\tt r.x.xu@pku.edu.cn}}

\begin{abstract}
The magnetar model and a solid quark star model for anomalous
X-ray pulsars/soft gamma-ray repeaters (AXPs/SGRs) are discussed.
Different manifestations of pulsar-like stars are speculated to be
due to both their nature (e.g., mass and strain) and their nurture
(ambience, and consequently the type of accretion) in the solid
quark star scenario.
Relevant arguments made by the author's group, including a debate
on solid cold quark matter, are briefly summarized too.
\end{abstract}


\end{frontmatter}

\section{Historical notes on magnetars and quark-stars}

Since the discovery of radio pulsars about 40 years ago, more
kinds of pulsar-like stars have been discovered, which are
distinguished by their various manifestations, i.e.,
accretion-driven X-ray pulsars, X-ray bursters, AXPs/SGRs
(anomalous X-ray pulsars/soft gamma-ray repeaters), CCOs (central
compact objects in supernova remnants), DTNs (dim thermal neutron
stars), and RRATS (rotating radio transient sources).
Among these, the enigmatic sources, AXPs and SGRs, share three
common surprising features: (i) all the sources have long spin
periods ($5-12$ s) and persistent X-ray emission with luminosities
well in excess of their spin-down powers, (ii) no clear evidence
for them in binaries is obtained, and (iii) they show X-ray
bursts, even giant flares saturating space detectors (the typical
luminosity is greater than $10^6$ times of the Eddington one).
Four SGRs (4 confirmed, 3 candidates) and eight AXPs (8 confirmed,
1 candidate) have now been discovered (see the catalog updated
online at
``http://www.physics.mcgill.ca/$\sim$pulsar/magnetar/main.html'').

Rapid rotation was generally thought to be the only energy source
for pulsar emission soon after the discovery of radio pulsars
\citep[e.g.,][]{mt77} until the discovery of accretion-powered
pulsars in X-ray binaries \citep{pr72}.
However, AXPs/SGRs have long spin periods (thus low spindown
power, their X-ray luminosities are much larger than their
spindown powers) and no binary companions, which rules out spin
and accretion in binary system as the power sources for the
emission.
It was then proposed that SGR-like bursts as well as the
persistent X-ray emission could plausibly be the result of field
decay of ultra-magnetic neutron stars if MHD dynamo action in the
proto-stars is very effective in case that the objects spin
initially at periods of $\sim 1$ ms \citep{DT92,TD93,TD95}.
Because of the starquakes in the crusts of normal neutron stars, a
self-induction electric field is created. The strong electric
field could initiate avalanches of pair creation in the
magnetosphere and certainly accelerate particles, resulting in a
so-called {\em magnetar corona} \citep{bt06}, from which
high-energy bursts could be observed. The power source of
AXPs/SGRs is actually the magnetic energy through
field-reconnection there \citep{wt05}. This magnetar model is very
popular nowadays in the astrophysical community.

Is there any other kind of possible power source for AXPs/SGRs?
This fundamental question, however, could be related to a much
more elementary and very important puzzle: {\em What is the real
nature of pulsar-like stars?} \cite[That is, are they really
neutron stars or quark stars? For a general review, see,
e.g.,][]{Horvath06,xu06a}.
In fact, alternative power sources were also proposed for
AXPs/SGRs in a solid quark star model \citep{Horvath05,xty06}.
Surely the properties of matter at high density but low
temperature have been of great interests in both theoretical and
observational aspects, especially in the case of compact stars
(see a discussion in \S4).
Quantum chromo-dynamics (QCD) is believed to be the underlying
theory for the elementary strong interaction, but its strong
coupling at low energy scales has troubled physicsists for a very
long time.
Lattice QCD simulations and effective QCD models are ways for us
to study the non-perturbative effects, whereas lattice QCD seems
not accessible for calculating supranuclear matter at low
temperature because of the still unsolved sign problem.
Therefore, physicists can only approximate the state of matter at
high density and low temperature with effective models.

As for the inner constitution of pulsar-like stars, nuclear matter
(related to {\em neutron stars}) is one of the speculations even
from the Landau's time, but quark matter (related to {\em quark
stars}) is an alternative due to the fact of QCD's asymptotic
freedom \citep{Bodmer71,Witten84,fj84}.
Recently, \cite{Jaikumar06} even proposed a crust comprised of
nuggets of strange quark matter embedded in an uniform electron
background, and found that a much reduced density gradient and
negligible electric field would exist in such a quark star.
It is a pity that we are now {\em not} able to determine
confidently which state exists in reality due to the difficulty of
QCD calculations. Therefore, we have to focus on astrophysical
observations in the research.
Based on the Planck-like spectrum without atomic features and the
precession properties of pulsar-like stars, a solid quark matter
conjecture was suggested \citep{xu03}. Additionally, other
features naturally explained within this model could possibly
include sub-pulse drifting in radio emission, glitches, strong
magnetic fields, birth after a successful core-collapse supernova,
detection of small bolometric radii, and transient bursts of GCRT
J1745$-$3009 \citep{xqz99,z04,xu05,yue06,zx06}.
Besides, the observational features of AXPs/SGRs may also reflect
the nature of solid quark stars, that will be focus on in this
paper.
Power sources in solid quark stars include the elastic energy, the
quake-released gravitational energy, and the accretion-released
gravitational energy of fossil disk \citep{xty06}.
The hard X-ray bursts of AXPs/SGRs are proposed to be the results
of AISq ({\em accretion-induced starquakes}) in the model.

\section{Possible issues challenging the magnetar idea}

Conventional models for AXPs/SGRs are magnetars, a kind of neutron
stars with surface magnetic fields in the range $\sim (10^{14} -
10^{15})$ G, especially for understanding three supergiant flares
(with released energy of $\sim 10^{44-47}$ erg) observed in SGRs.
The quiescent X-ray emission with luminosity $\sim 10^{34 - 36}$
erg/s as well as the bursts of AXPs/SGRs are supposed to be
powered by magnetic field decay.
However, there are still some observational or theoretical
arguments which may challenge the idea, although magnetars are
really popular in the astrophysical community nowadays.

In fact, it is still debatable to estimate the magnetic fields
($\propto \sqrt{P{\dot P}}$) of AXPs/SGRs simply from their period
($P$) and period derivative ($\dot P$), since AXPs/SGRs could not
be braked dominantly by the magnetodipole torque.
AXPs/SGRs were suggested to be normal-field pulsar-like stars
which are in an accretion propeller phase \citep{Alpar01,chn00}.
But the difficulty in the latter view point is to reproduce the
irregular bursts, even the superflares. That difficulty could be
overcome if bursts are proposed to be the results of starquakes of
solid quark stars \citep{owen05,xty06}.
In addition, some problems with the magnetar scenario may arise
from recent relevant research, which are summarized as following.

(i). The superstrong fields of magnetars are supposed to be
created by MHD-dynamo action of rapid rotating protoneutron stars
with spin period $<3$ ms. The Poynting flux and the relativistic
particle ejection of such a star should power effectively the
supernova remnants.
Numerical simulations of supernova remnant expansion with internal
magnetars did shown a faster expansion when the energy injected
into the supernova remnant by magnetar spin-down is taken into
account \citep{ah04}.
Such energetic remnants are then expected in magnetar models, but
have not been detected \citep{vk06}.

(ii). Dust emission around pulsar-like stars (e.g., AXPs) was
proposed to test observationally the propeller scenarios of quark
stars with Spitzer or SCUBA \citep{xu05,xu06c}. Actually a recent
discovery of mid-infrared emission from a cool disk around an
isolated young anomalous X-ray pulsar, 4U 0142+61, has been
reported \citep{wck06} although it is still a matter of debate
whether significant propeller torque of fallback matter acts on
the star.

(iii). In order to understand the thermal Planck-like spectra in
the conventional model of neutron star atmospheres for DTNs or
CCOs, strong magnetic fields (and possibly rapid rotation) are
suggested.
However, a problem inherent in these efforts is \citep{xu04}: {\em
How to calm down the magnetospheric activities so that no
AXP/SGR-like persistent and bursting emission could be radiated?}
This problem exists also for young ``high-field'' radio pulsars
since no magnetar-like emission is detected from them
\citep{1119}.
The facts could be understandable if both AXPs/SGRs and
``high''-field pulsars are braked effectively by propeller torques
of fossil disks, with the relative accretion rates (being
indicated by the ratio of co-rotation radius to magnetospheric
one) to be higher in the former than in the latter.

(iv). It is still a matter of debate whether the absorption
features in SGR 1806-20 can be interpreted by proton or electron
cyclotron resonance \citep{xwq03}. The field is only $\sim 5\times
10^{11}$ G in the context of electron cyclotron lines.

(v). The pressure should be very anisotropic in a relativistic
degenerate neutron gas in equilibrium with a background of
electrons and protons when the magnetic field is stronger than the
critical field ($\sim 10^{13}$ G). The vanishing of the equatorial
pressure of the gas might result in a transverse collapse, and a
stable magnetar could then be unlikely \citep{mrs03}.
Such kind of quantum-magnetically induced collapse could also be
possible in quark stars with ultra strong magnetic fields
\citep[$>10^{\sim 19}$ G,][]{mrs05}.
In this sense, quark matter would tolerate much higher magnetic
field than neutron matter, and no neutron star with $10^{\sim
(14-15)}$ (i.e., {\em no} magnetar) could form stably.
Nevertheless, more theoretical investigations of the equilibrium
of nuclear or quark matter in general relativity are necessary.

In addition, the implied transverse velocity of an AXP, XTE
J1810-197, indicates that its kick velocity is completely ordinary
($<400$ km/s) though the AXP manifests strikingly
\citep{Helfand2007}. It might be not easy to reproduce normal
kicks for neutron stars with very fast rotation (initial period
$P<3$ ms) and much high magnetic fields ($B\gg 10^{13}$ G) in the
magnetar model, since the initial magnetic field and the initial
spin are generally considered to be key parameters for kicks
\citep{Lai2004}.
Considering these criticisms, we think that alternative ideas for
understanding AXPs/SGRs are welcome.
We note that the quark-star model could be competitive, in which
both gravitational and elastic energies would power the photon and
particle radiation of AXPs/SGRs.

\section{AXPs/SGRs: quark-stars with fossil disks?}

Why do pulsar-like stars behave so diversely?
As noted by \cite{xu06c}, two factors could be very important for
their different manifestations: {\em environment} and {\em stellar
mass}.
A dense environment may be responsible for leaving a
remnant-nebula after a supernova and for resulting possibly in a
fallback accretion (e.g., via disk) around the compact star,
whereas starquakes could  more probably occur in massive compact
stars. Bursts, as the results of quakes, could frequently appear
in an accreting solid quark star (even low-mass) because of AISq
(accretion-induced starquakes).
In this context, various manifestations of pulsar-like stars would
be speculated to be related to quark stars with different nature
and nurture (fossil disks, quakes, and masses; see Table 1).
\begin{table}[]
\caption[]{Speculated characters of pulsar-like stars in the quark star scenario.}
  \label{Tab:publ-works}
  \begin{center}\begin{tabular}{c|c|c|c}
  \hline\noalign{\smallskip}
  \hline\noalign{\smallskip}
  {\bf manifestation} &  {\bf propeller torque} & {\bf starquake} & {\bf mass}\\
  \hline\noalign{\smallskip}
\parbox[t]{.7in}{\raggedright ~\\Radio\\pulsars\\ ~\\}\vline{}
\parbox[t]{.9in}{\raggedleft {\em normal}\\{\em millisecond}\\
{\em part-time}\\{\em ``high''-field}} &
\parbox[t]{.8in}{\raggedright negligible\\negligible\\fossil disk?\\significant}&
\parbox[t]{.8in}{\raggedright glitches\\microglitch$^*$\\~\\glitches}&
\parbox[t]{.9in}{\raggedright some low?\\most low?\\low (dying)?\\~}\\
  \hline\noalign{\smallskip}
\parbox[t]{.7in}{\raggedright Accretion-\\powered}\vline{}
\parbox[t]{.9in}{\raggedleft {\em X-ray pulsars}\\{\em X-ray bursts}} &
\parbox[t]{1.1in}{\raggedright strong accretion\\strong accretion}  &
a frequency glitch$^\dag$ & \\
 \hline\noalign{\smallskip}
CCOs  & significant & no-detection  & low?\\
\hline\noalign{\smallskip}
DTNs &  weak accretion & no-detection  & low?\\
\hline\noalign{\smallskip} AXPs & substantial &  AISq$^\ddag$, frequent$^\flat$  & \\
\hline\noalign{\smallskip} SGRs  & substantial & AISq, more
frequent &
 \parbox[t]{1.1in}{\raggedright$\sim M_\odot$ for those\\
with superflares}\\
    \hline\noalign{\smallskip}
    \hline\noalign{\smallskip}
\end{tabular} \end{center}
$^*$ Only one microglitch with $10^{-11}$ change in fractional
frequency was detected in a millisecond pulsar, PSR B1821$-$24
\citep{cb04}.\\
$^\dag$ Still one glitch in an accreting pulsar, KS 1947+300, has
been reported during outbursts, with a fractional change in
frequency of $3.7\times 10^{-5}$ to be significantly larger than
the values observed in the glitches of radio pulsars
\citep{gml04}.\\
$^\ddag$ Accretion-induced starquakes (AISq).\\
$^\flat$ Large glitches are detected from two AXPs, nearly
simultaneously with burst: 1E 2259+586 ($\Delta\nu/\nu\sim 4\times
10^{-6}$) and CXOU J164710.2-455216 ($\Delta\nu/\nu = 6.5\times
10^{-5}$) \citep{Israel07}.
\end{table}
According to the calculation of Fig. 2 in \cite{xty06}, more
gravitational energy could be released during quakes of solid
quark stars with higher masses, and we may expect that SGRs
emitting giant flares could be quark stars with $\sim
(1-2)M_\odot$.

By assuming that AXPs/SGRs and ``high''-field radio pulsars could
be torqued effectively by fossil disks, \cite{yzx06} suggest that
SGR 1900+14 and AXP 1E 1048.1-5937 might be quark star with $\sim
10^{-(1-2)}M_\odot$ mass if all AXPs and SGRs are likely grouped
together in their Fig. 1.
One of the ``high''-field pulsar, PSR J1846-0258 in the supernova
remnant Kes 75, has only been detected in the X-ray band, with a
much higher ratio of X-ray luminosity to the spindown energy loss
than that found in other young radio pulsars (e.g., the Crab
pulsar). The timing of PSR J1846-0258 behaves like that of radio
pulsars, with a braking index of 2.65 \citep{Livingstone06}.
Though the emission luminosity is smaller than the spindown power,
it is still possible that PSR J1846-0258 would be actually
accretion-powered if the real magnetic moment is very low.
In this case, the monopole-induced potentials drop in the open
field region of PSR J1846-0258 might not be high enough (due to a
much lower real magnetic field) to produce significant pair plasma
for radio emission.
Also the potential drop of AXPs/SGRs could be too low to initiate
avalanches of $e^\pm$ pairs in their quiescent states. However, it
is worth noting that radio activity would still be possible for
AXPs/SGRs after outburst, as in the case of radio emission from
XTE J1810$-$197 \citep{Camilo06}, since strong pair plasma could
form due to quake-induced $E_\parallel$ (electric field parallel
to magnetic field).
In a conclusion, AXPs and ``high''-field radio pulsars might be in
a same group, with different magnetic moment but similar
appearances.

The magnetospheric electrodynamics of isolated rotation-powered
pulsars is studied extensively in the literatures, including pair
production, particle acceleration, and $E_\parallel$-solutions.
However, the corresponding dynamics of a rotating quark star in a
propeller phase, with significant accretion onto the polar cap,
could be in a very different scenario.
Ruderman-Sutherland-type vacuum gap \citep{rs75} may not exist
(thus no sparking occurs) above polar caps since the hot polar cap
bombarded by accreted nucleons could emit sufficient charged
mesons and/or hadrons in the QCD-phase conversion process.
A space charge-limited flow model \citep{as79} may adaptable in
this case, where the mesons and hadrons are accelerated
effectively by $E_\parallel$ and may annihilate finally (or do by
other interactions) into ultra-high energy photons and neutrinos
in magnetospheres.
TeV-photon emission observed could be understood in this picture,
which has also been noted by many authors
\citep[e.g.,][]{zd03,lb05,nc06}. Certainly, the related dynamics
needs further investigations in order to study the detail
astrophysics there.

We note that fossil disk could affect the spindown torque in two
aspects. (i) Propeller torque attributes to the MHD-coupling
between accretion flow and co-rotating particles at the
magnetospheric radius ($r_{\rm m}$). (ii) Particles (accelerated
by $E_\parallel$) with higher number density would be ejected (and
the star spins down faster) if $r_{\rm m}$ is smaller than the
light cylinder radius ($r_{\rm lc}$; the open field line region is
then determined by $r_{\rm m}$ rather by $r_{\rm lc}$). In this
case, the potential drop in the open field lines is higher than
that expected for pure rotation-powered pulsars.

{\em Conclusive evidence for quark-stars?} \cite{xu06c} had
addressed that following observations could test for the quark
star (even with low mass) idea:
(i) to detect dust emission around pulsar-like stars; (ii) to
geometrically determine the radii of distant pulsar-like stars by
advanced X-ray or UV-band facilities; (iii) to detect
gravitational waves from pulsar-like stars; and (iv) to search
sub-millisecond pulsars (especially with periods $<0.5$ ms).
Although it should certainly be very clear evidence for quark
stars if we could detect sub-millisecond pulsars by advanced radio
telescope (e.g., the FAST) in the future \citep{xu06d}, we note
that to detect dust emission form pulsar-like stars is another way
to find such evidence.
The maximum elastic quadrupole deformations sustainable by normal
neutron stars and solid strange quark stars might result in
distinguishable maximum ellipticities~\citep{owen05}, and the
gravitational radiation may show the nature of pulsar-like stars.
Though we shouldn't detect gravitational wave because of the low
spindown powers in the second LIGO science run
\citep{LIGO05,owen06}, a result of search for known pulsars in
future LIGO run might only show upper limits of masses (or radii)
for the targets since the gravitational wave radiation could be
mass-dependent \citep{xu06b}.
It is also worth noting that gravitational wave emission
associated with the energetic superflares of SGRs may be gathered
from the LIGO data \citep{Horvath05},
and that, in fact, \cite{ligo2007} had put an upper limit on the
gravitational wave emission, $7.7\times 10^{46}$ erg, for the 92.5
Hz QPO of SGR 1806-20.
In addition, it is still not sure whether the small radii derived
from Planckian thermal radiation indicate the real stellar
surfaces or only the hot polar caps, and it is then very necessary
and important to observe confidently the emission from global
surface of pulsar-like stars in, e.g., soft X-ray or UV bands.
To find a binary system of low mass quark star (formed via AIC)
and white dwarf could also be expected in the quark star model,
which might be an effective way to search quark stars with very
low masses.

\section{Conclusions and Discussions}

One of the daunting challenges nowadays is to understand the
fundamental strong interaction between quarks, especially in the
low-energy limit, since the coupling is asymptotically free in the
limit of high-energy.
Though quark matter is predicted in QCD because of asymptotic
freedom, astrophysical quark stars would be still in a low-energy
region where quarks are coupled non-linearly.
In this sense, AXPs/SGRs (including other pulsar-like stars) would
be ideal laboratories for us to study the non-perturbative effects
of QCD.
Phenomenologically, a solid state of cold quark matter was
suggested, on which understanding the peculiar features of
AXPs/SGRs could be based. It is suggested in this paper that the
various manifestations of pulsar-like stars could imply different
environments and inner structure of the sources, and that the
bursts of AXPs/SGRs could be the results of accretion-induced
starquakes (AISq) of quark stars in their propeller phases.
A solid quark star quakes as stress energy develops due to
spin-down, or cooling, or even general relativistic effect.

There exist other versions \citep[e.g.,][]{oed04,ouyedI,ouyedII}
of quark star models for AXPs/SGRs, which may show the necessary
of introducing astrophysical quark matter for understanding this
kind of sources.
Assuming a superconducting state, \cite{Niebergal06} calculated
the magnetic field decay due to the expulsion of spin-induced
vortices, and found that the magnetic field strengths and periods
remain almost unchanged for typical parameters.
Though glitches in the model of Ouyed et al. could be explained
with intricate details, glitches and quasi-periodic oscillations
(QPOs) during superflares might hint that the cold quark matter
could be in a solid state.
A simple toy-model \citep{gsa06}, with the inclusion of global
MHD-coupling between the elastic crust and the fluid core of a
normal neutron star, was suggested to explain QPOs of SGRs, but a
similar model in the solid quark star scenario is worth to make in
order to understand the detail observations and to know
observationally the features of solid quark matter (e.g., the
shear modulus), since MHD waves in the global magnetospheres could
be excited by oscillations of quark stars after their quakes.
It is also very necessary to calculate the quake-induced
$E_\parallel$ (electric field parallel to magnetic field) and
particle acceleration in both open and closed field line regions
in order to model the magnetospheric activity of AXPs/SGRs after a
starquake.
Two large glitches ($\Delta\nu/\nu\sim (10^{-5}-10^{-6})$) are
detected in AXPs \citep{Israel07,muno2007}, and it would be very
urgent to search glitches in SGRs since larger glitches could be
natural in this AISq scenario.

Actually, from a viewpoint of condensed matter physics, symmetries
break as energy-scale decreases so that different states of matter
are in the Universe.
Cold quark matter was supposed to be approximated by ideal Fermi
gas when the interaction between quarks is neglected for matter to
be asymptotically dense. But a superfluid state, in which the
global U(1) symmetry is broken, was also proposed since any
attractive interaction between quarks may render the Fermi surface
unstable.
It is further suggested that a crystalline structure (i.e., the
translation symmetry is broken) may form in this color
superconductivity quark matter, and that the shear modulus could
be 20 to 1000 times larger than those of neutron star crusts
\citep{mrs07}.
However, another possibility of condensation in position space
could still not be rule out. Quark clusters would form in this
scenario, and quark matter would be solidified at low temperature.
An investigation on the initial spin-evolution may show signatures
of quark star solidification. Star's cooling and gravitational
radiation would affect its viscosity and rotation.

{\em From sQGP to solid quark matter?}
The non-linear complexity of strong interaction at low energy
could lead to two hot points in recent researches: the physics of
relativistic heavy ion collision and the study of pulsar-like
star's structure.
A key point in the quark-star model suggested for AXPs/SGRs is the
conjecture that quarks would be clustered in strongly coupled
quark-gluon plasma (sQGP), especially in the supranuclear matter
with low temperature \citep{xu03,xu04,xu05,xu05b} when the
interaction parameter $\Gamma$ (defined by the ratio of the
potential energy to the kinetic energy) is very high. {\em Solid}
quark matter could be very likely in case that the thermal
kinematic energy of quark clusters is much lower than the
interaction energy between the clusters (i.e., $\Gamma\gg 1$).
In fact, it is recently realized $\Gamma$-value of the QGP at RHIC
should be much higher than that predicted by the perturbative
theory of QCD~\citep[e.g.,][]{Shuryak}. Therefore, the QGP created
at RHIC is also sQGP, and one may expect an even higher $\Gamma$
for cold quark matter (i.e., temperature to be much lower than 1
MeV). Glass-like or crystal-like solid quark matter would then be
possible in case of keV-temperature.
One way to test the phenomenological interaction-models for sQGP
at laboratory could be calculating the quark clusters by qMD
(quark molecular dynamics), where the colored and flavored quarks
could be treated as semi-classical particles interacting via a
Cornell-potential~\citep{Hofmann},
$$%
V(r) = -\frac{3}{4}\frac{\alpha_{\rm s}}{r}+\kappa r.
$$%
Strong interaction via this potential may favor formation of quark
clusters, although the interacting potential between quark
clusters could be in a different type.
The experimental yield of multi-quark clusters (pentaquark,
teraquark) may improve our understanding of sQGP.
Recent achievements about classical sQGP are presented
in~\cite{GelmanI,GelmanII}.
Therefore, we expect that the idea of quark clustering in cold
quark matter would probably be tested experimentally.

{\em The beauty of non-perturbative QCD?} The uncertainty of
determining the composition of pulsar-like stars depends actually
and strongly on the non-perturbative nature of QCD at low-energy
level, which is unfortunately not well understood. This is the
real reason that we {\em cannot} make sure pulsars are neutron or
quark stars according only to today's QCD calculations.
Pessimistically, it might be impossible {\em forever} to know the
state of supranuclear matter by pure QCD calculations; and we
should then have to recognize the importance of observations and
phenomenology in studying pulsar's inner structures.
%
Investigations in such a way may discover the natural ``beauty''
of non-perturbative QCD.
%
As in another but simple kind of non-perturbative dynamics, the
non-linear phenomena in fluid dynamics (such as turbulence and
chaos), which are surely not able to be determined by the famous
Navier-Stokes equation, do exist in nature and are beautiful.
In this sense, it is not against the rule of nature's beauty to
study observationally and phenomenologically the elementary strong
interaction at low energy.

{\em Acknowledgments}.
I thank Jorge E. Horvath, Rachid Ouyed, and Benjamin J. Owen for
their helpful suggestions and useful comments on the science as
well as language of the manuscript. Two referees are also
sincerely acknowledged for their constructive suggestions.
This work is supported by National NSF of China (10573002,
10778611) and by the Key Grant Project of the Chinese Ministry of
Education (305001). I would like to thank Youling Yue and other
members in the pulsar group of Peking University for their
stimulating discussions.

\end{document}